\documentclass[12pt]{iopart}

\usepackage{graphicx}
\usepackage{iopams}
\begin{document}
\title{Interplay of diffraction and nonlinear effects in propagation of ultrashort pulses}

\author{C L Korpa$^1$, Gy T\'oth$^2$ and J Hebling$^{1,2}$}

\address{$^1$ Institute of Physics, University of P\'ecs, Ifj\'us\'ag \'utja 6, 7624 P\'ecs, Hungary}
\address{$^2$ MTA-PTE High-Field Terahertz Research Group, Ifj\'us\'ag \'utja 6, 7624 P\'ecs, Hungary}
\eads{korpa@fizika.ttk.pte.hu, tothgy@fizika.ttk.pte.hu, hebling@fizika.ttk.pte.hu}

\begin{abstract}
We investigate the interplay of diffraction and nonlinear effects during propagation of very short 
light pulses. Adapting the factorization approach to the problem at hand by keeping the transverse-derivative
terms apart from the residual nonlinear contributions we derive an unidirectional propagation equation valid for weak dispersion and reducing to the 
slowly-evolving-wave-approximation for the case of paraxial rays. Comparison of numerical simulation 
results for the two equations shows pronounced differences when self-focusing plays important role. 
We devote special attention to modelling propagation of ultrashort terahertz pulses taking into account diffraction as well as Kerr type and
second order nonlinearities.
Comparing measured and simulated wave forms we deduce the value of the nonlinear refractive index of lithium niobate in the terahertz region to be
three orders of magnitude larger than in the visible. 
\end{abstract}


\vspace{2pc}

\submitto{\JPB}
\maketitle

\section{Introduction}

Improving theoretical description of propagation of light pulses consisting of just a few cycles has received 
considerable interest in the past two decades \cite{BK-1997,Gaeta-1998,Porras-1999,Gaeta-2002,Tara-2001,Kolesik-2002,
Kolesik-2004,Kinsler-2003,Kinsler-2005,Kinsler-2010,Kinsler-2010a}. However, relatively small number 
of investigations consider the variation in the transverse directions i.e.\ the diffraction effects. 
For example, analytical solutions are presented in \cite{Porras-1999} for diffraction and dispersion effects but only for
the case of linear propagation. Our aim is to consider both nonlinear effects and transverse variation leading
to broadening in the transverse direction but also self focusing for high intensities for the case when dispersion effects 
including attenuation have a minor role. 

Introducing additional dependence on one or both transverse coordinates
makes mathematical modelling and numerical simulations more involved but is necessary for studying effects like self-guiding
and filament formation when competition between diffraction and nonlinearity plays important role 
\cite{Kovachev-2007,Belyaeva-2011,Kovachev-2013,Kovachev-2013a,Kovachev-2014}. Nonlinear envelope equations of the type 
derived in \cite{BK-1997} were used for numerical simulation of shape effects \cite{Kovachev-2007} on the propagation of femtosecond pulses in media
with weak dispersion and for studying filament stability \cite{Kovachev-2013}. Analytical approach using soliton solutions can also be used for
filament propagation description \cite{Kovachev-2014} and collision-dynamics study of otical pulses \cite{Kovachev-2013a} for certain fixed 
values of the nonlinear coefficient. Taking into account both the transverse (as well as longitudinal) profile and nonlinearity is essential in
describing interaction of optical and matter-wave soliton solutions characterizing trapped-atom arrangements \cite{Belyaeva-2011}.

Examination of the term containing the transverse derivatives reveals that in the traditional slowly-varying-envelope 
approximation an implicit expansion in the transverse components of the 
wave vector is made by assuming that they are much smaller than the longitudinal component pointing in the propagation 
direction \cite{Rothenberg-1992}. This approximate treatment of transverse dependence also characterizes for example the 
slowly-evolving-wave approximation (SEWA) introduced in \cite{BK-1997} as pointed  out in \cite{Kolesik-2002}. 
We avoid this expansion pertaining to paraxial-ray approximation by modifying the factorization approach of
Kinsler \cite{Kinsler-2010} to obtain an evolution equation suitable for numerical calculation which properly takes into 
account diffraction including space-time focusing effects.

\section{Theoretical considerations}
\label{s1}

To obtain an unidirectional propagation equation we follow the approach outlined in \cite{Kinsler-2010} and 
discussed earlier in more detail in \cite{Ferrando-2005}. 
We represent the physical electric field ${\bi E}_{ph}({\bi r},t)$ by analytic complex field ${\bi E}({\bi r},t)$ 
whose Fourier transform contains only positive frequencies \cite{Haykin-2001,Conforti-2010}:
\begin{equation}
{\bi E}_{ph}({\bi r},t)=\frac{1}{2}\left[ {\bi E}({\bi r},t)+\mbox{c.c.}\right].
\end{equation}
Taking the propagation direction to be the $z$ axis
the 3-dimensional wave eqation can be written as
\begin{eqnarray}
\left( \partial_z^2+\nabla_\perp^2 \right){\bi E}({\bi r},t)&-&\frac{1}{c^2}\frac{\partial^2}{\partial t^2}\epsilon_L(t)\star {\bi E}({\bi r},t)=
\frac{4\pi}{c^2}\frac{\partial^2}{\partial t^2}{\bi P}_{NL}({\bi E},{\bi r},t)\nonumber \\ 
 &+&\mu_0 \frac{\partial {\bi J}({\bi r},t)}{\partial t}+\bi\nabla \, \bi\nabla\cdot {\bi E}({\bi r},t) \label{eq1}
\end{eqnarray}
where $\star$ denotes convolution, $\epsilon_L$ contains the linear part of the material response, ${\bi P}_{NL}({\bi E},{\bi r},t)$ 
is the nonlinear polarization, ${\bi J}({\bi r},t)$ the current density and we assumed a non-magnetic material. Transforming into
wave-vector and frequency space we can write (\ref{eq1}) as
\begin{equation}
\left( -k_z^2-k_\perp ^2+\beta^2\right){\bi E}({\bi k},\omega)=-{\bi Q}, \label{eq2}
\end{equation}
with $-{\bi Q}$ denoting all terms corresponding to the right-side of (\ref{eq1}) and $\beta^2=\omega^2 \epsilon_L(\omega)/c^2$.
Note that in contrast to \cite{Kinsler-2010} we do not place the term with transverse derivatives in the residual term ${\bi Q}$
but keep it exlicitly. Performing the factorization in (\ref{eq2}):
\begin{equation}
\left( k_z-\sqrt{\beta^2-k_\perp^2}\right)\left( k_z+\sqrt{\beta^2-k_\perp^2}\right){\bi E}({\bi k},\omega)={\bi Q}, \label{eq3}
\end{equation}
we can express ${\bi E}$ as a sum of two terms:
\begin{equation}
{\bi E}({\bi k},\omega)=\frac{1}{2\sqrt{\beta^2-k_\perp^2}}\left( \frac{1}{k_z-\sqrt{\beta^2-k_\perp^2}}-\frac{1}
{k_z+\sqrt{\beta^2-k_\perp^2}}\right) {\bi Q}.
\end{equation}
Defining ${\bi E}^{(\pm)}$ by
\begin{equation}
{\bi E}^{(\pm)}({\bi k},\omega)=\frac{\pm 1}{2\sqrt{\beta^2-k_\perp^2}}\,\frac{1}{k_z\mp \sqrt{\beta^2-k_\perp^2}}{\bi Q}
\end{equation}
we can write ${\bi E}={\bi E}^{(+)}+{\bi E}^{(-)}$ where ${\bi E}^{(\pm)}$ obey the equations:
\begin{equation}
\left( k_z\mp \sqrt{\beta^2-k_\perp^2} \right) {\bi E}^{(\pm)}({\bi k},\omega)=\frac{\pm 1}{2\sqrt{\beta^2-k_\perp^2}}{\bi Q}. \label{res0}
\end{equation}
Transforming back to ordinary space with respect to $k_z$ (but not ${\bi k}_\perp$) we obtain the two equations governing
propagation of ${\bi E}^{(\pm)}(z,{\bi k}_\perp,\omega)$:
\begin{equation}
\partial_z {\bi E}^{(\pm)}(z,{\bi k}_\perp,\omega)=\pm \rmi \sqrt{\beta^2-k_\perp^2} {\bi E}^{(\pm)}(z,{\bi k}_\perp,\omega)\pm 
\frac{\rmi }{2\sqrt{\beta^2-k_\perp^2}}\,{\bi Q}. \label{res1}
\end{equation}
Equations (\ref{res1}) can be regarded as an alternative derivation of the unidirectional pulse propagation equation of 
Kolesik and Moloney \cite{Kolesik-2004} which is based on modal expansion of the fields. A potential disadvantage 
of relying on result (\ref{res1}) is that one
has to work in the perpendicular wave vector space and without expansion one can not meaningfully transform to 
perpendicular derivatives in the ordinary space. Since our final intention is to arrive at an unidirectional propagation
equation using only the forward propagating field ${\bi E}^{(+)}$ we have to mention an important constraint arising from
attenuation taken into account through the (positive) imaginary part of $\beta^2(\omega)$. That term provides suppression for the 
forward propagating component but as seen form (\ref{res1}) it enhances the backward propagating term. Poor phase
matching is expected to provide suppression for backward propagation but for long enough propagation distances compared to
attenuation length that component may be amplified enough in order to make its effect through nonlinearity non-negligible.
As a consequence, for the applicability of the unidirectional equation we require the condition of weak dispersion to be 
satisfied for the full frequency spectrum of the 
pulse and the propagation distance to be shorter than the attenuation length. This is in accordance with constraints on the
applicability of the slowly-evolving-wave-approximation \cite{BK-1997} concerning weak dispersion and restricted propagation length
as discussed in \cite{Xiao-1999} and section 11.6 of \cite{Oughstun-2009b}. 

For a general transverse dependence one can use the cartesian components of the transverse wave vector $k_x,k_y$ and
the corresponding eigenfunctions $\exp(ik_x x+ik_y y)$ as in \cite{Kolesik-2004}. However, since our
main interest resides in cylindrically symmetric pulse profiles we exploit that symmetry to make the numerical
modelling less computationally intensive and use the simulation with arbitrary transverse dependence only as a check of
numerical procedures. 

We make the transformation to transverse wave vector space using the fact that the solution of the eigenvalue problem
\begin{equation}
\nabla_\perp^2 f(r,\phi)+k_\perp^2 f(r,\phi)=0,
\end{equation}
with $r,\phi,z$ being the cylindrical coordinates, is \cite{Agrawal-2013}:
\begin{equation}
f(r,\phi)=\sum_{m=0}^\infty \left[ A_m\,J_m(k_\perp r)+B_m\,N_m(k_\perp r)\right] e^{i m \phi},
\end{equation}
where $J_m$ is the Bessel function and $N_m$ is the Neumann function of order $m$. 
Assuming uniform medium in the transverse direction and regularity of the solution at $r=0$ with rotational symmetry
around the $z$ axis means that only the $J_0$ term survives. We can thus expand the transverse dependence as
\begin{equation}
f(r)=\int_0^\infty A(k_\perp)\,J_0(k_\perp r)\rmd k_\perp \label{exptr1}
\end{equation}
and using the integral relation
\begin{equation}
\int_0^\infty r J_0(k'_\perp r)\,J_0(k_\perp r)\rmd r=\frac{1}{k_\perp}\delta(k'_\perp-k_\perp),
\end{equation}
where $\delta(k)$ is the Dirac delta function, we can invert (\ref{exptr1}) to get
\begin{equation}
A(k_\perp)=k_\perp \int_0^\infty r f(r) J_0(k_\perp r) \rmd r.
\end{equation}

There is an alternative approach to expression (\ref{exptr1}) which relies on explicit compactification 
of the transverse coordinate. Even without a strict boundary condition that the electric field vanishes 
at a fixed $r$ value, we may require that it be zero at some (large) value of $r=a$. That condition
restricts the continuous integration variable in (\ref{exptr1}) to discreet values $k_m^\perp$
satisfying the condition $J_0(k_m^\perp a)=0$ and thus leads to 
\begin{equation}
f(r)=\sum_{m=1}^\infty A_m\,J_0(k_m^\perp r). \label{exptr2}
\end{equation}
Denoting the zeros of the Bessel function $J_0(x)$ by $x_m,\;m=1,2,3,\ldots$ we can write $k_m^\perp=x_m/a$.
Using the orthogonality relation \cite{Jackson-1999}
\begin{equation}
\int_0^a r J_0(x_m r/a)\,J_0(x_n r/a) \rmd r=\frac{a^2}{2}J_1(x_n)^2 \,\delta_{nm}
\end{equation}
one can compute the expansion coefficients:
\begin{equation}
A_m=\frac{2}{a^2 J_1(x_m)^2}\,\int_0^a r f(r) J_0(x_m r/a) \rmd r.
\end{equation}
We used this approach as a check of numerical implementation based on (\ref{exptr1}) and indeed ascertained that 
in our numerical simulations the above two approaches give indistinguishable results if the value $a$ of the
radius at which the field vanishes is chosen large enough and for considered limited propagation distances. 
We acknowledge that the above compactification 
should be regarded only as an approximate numerical procedure and used with care since it leads to violation of   
general integral properties \cite{Lalor-1968,Sherman-1968,Sherman-1969} of angular-spectrum representation for diffracted wave fields. 

If the residual term on the right side of  (\ref{res0}) does not mix the forward propagating field
${\bi E}^{(+)}$ with the backward propagating ${\bi E}^{(-)}$ considerably, i.e.\ it is small compared
to propagation terms not mixing the two field components then one can decouple the corresponding 
equations for forward and backward propagation.

When taking into account both forward and backward propagating fields introduction of envelopes 
with slower spatial (in the $z$ direction) variation does not bring any real advantage 
since that introduces fast varying contributions in the residual term which mixes the two fields. 

If the mixing of forward and backward propagating fields through the residual term is small enough 
so that a unidirectional propagation applies to good accuracy it is advantageous to introduce the
envelope 
${\bi A}(z,{\bi r}_\perp,t)$ with slower $z$ and $t$ variation than the (complex) field itself through
\begin{equation}
{\bi E}^{(+)}(z,{\bi r}_\perp,t)={\bi A}(z,{\bi r}_\perp,t)\, \exp[\rmi (\beta_0 z-\omega_0 t)],
\end{equation}
where $\omega_0$ is the predetermined central (carrier) frequency and $\beta_0=n(\omega_0)\omega_0/c$. In this way
 (\ref{res1}) for the forward propagating field is replaced by the following expression:
\begin{eqnarray}
\partial_z {\bi A}(z,{\bi k}_\perp,\omega)& = & \rmi (\sqrt{\beta^2-k_\perp^2}-\beta_0) {\bi A}(z,{\bi k}_\perp,\omega) \nonumber \\
&+&
\frac{\rmi }{2\sqrt{\beta^2-k_\perp^2}}\,{\bi Q}_A({\bi A}), \label{ueq1}
\end{eqnarray}
with the residual term on the right side depending also on the envelope. It is customary to introduce the
moving frame by transforming to new time $\tau$ and longitudinal distance $\zeta$ variables \cite{BK-1997}:
\begin{equation}
\tau=t-\beta_1 z,\; \zeta=z,
\end{equation}
with $v_g(\omega_0)=1/\beta_1$ the group velocity at $\omega_0$. The unidirectional envelope evolution equation 
(\ref{ueq1}) then becomes
\begin{eqnarray}
\partial_\zeta {\bi A}(\zeta,{\bi k}_\perp,\omega)&=& \rmi \left[ \sqrt{\beta^2-k_\perp^2}-\beta_0-\beta_1 \omega \right]
 {\bi A}(\zeta,{\bi k}_\perp,\omega)\nonumber \\
&+&\frac{\rmi }{2\sqrt{\beta^2-k_\perp^2}}\,{\bi Q}_A({\bi A}). \label{ueq}
\end{eqnarray}

\section{Numerical simulation}
\label{s2}

We now turn to numerical simulation of light pulse propagation. As a first example we consider the case 
examined in \cite{BK-1997} but without diffraction effects being taken into account there. A Kerr type nonlinearity is assumed with
electric field and nonlinear polarization having the same direction. This is a reasonable approximation since
the nonlinear effect in itself is small and including a small correction in the form of projection on the electric field vector 
is not expected to be significant \cite{Kolesik-2004}. In our simulations we do distinguish the transverse and longitudinal
components of the electric field with respect to the axis of propagation based on values of the length of wave vector and its 
transverse component and add up these two components of electric field separately. However, for the cases studied numerically 
in the following we observe that this 
separation of transverse and longitudinal components gives noticeable difference compared to the case when neglecting it only for
large enough transverse-coordinate values where typically the electric field magnitude drops to 1--2\% of its central
value. After transforming to $\omega$ (i.e.\ $\beta(\omega)$) and $k_\perp$ space the scalar form of (\ref{ueq}) is thus
appropriate:
\begin{eqnarray}
\partial_\zeta A(\zeta,{\bi k}_\perp,\omega)&=& \rmi \left[ \sqrt{\beta^2-k_\perp^2}-\beta_0-\beta_1 \omega \right]
 A(\zeta,{\bi k}_\perp,\omega)\nonumber \\
&+&\frac{2\pi \rmi (\omega+\omega_0)^2}{c^2\sqrt{\beta^2-k_\perp^2}}\, B(A), \label{ueqs}
\end{eqnarray} 
where the nonlinear polarization has been written as 
\[
P_{NL}(z,{\bi r}_\perp,t)=A_{NL}(z,{\bi r}_\perp,t,A)\,\exp[\rmi (\beta_0 z-\omega_0 t)]+\mbox{c.c.},
\]
and $B(A)$ is the Fourier transform of $A_{NL}$ to frequency and perpendicular wave vector space. If we expand
$\sqrt{\beta^2-k_\perp^2}$ in the first term of (\ref{ueqs}) to linear order in $k^2_\perp$ and in the 
term containing $k^2_\perp$ approximate the refractive index $n(\omega)$ with its value at the carrier frequency $\omega_0$
we recover the linear terms of the SEWA equation (6) in  \cite{BK-1997}.
We also recover the nonlinear term in that equation if in our nonlinear term we put $k_\perp$ to zero and again 
take the refractive index frequency independent and equal to its value at $\omega_0$. 
Simulation results show that these simplifications are quite good 
approximations for the case considered in \cite{BK-1997} consisting of a pulse with central 
wavelength $\lambda_0=0.8\;\mu$m propagating in fused silica with a hyperbolic-secant shaped envelope
$A_s(t)=1/\cosh\,[1.76 t/\tau_p]$ and $\tau_p=2.67\;$fs. However, pronounced difference characterizes the solutions of the 
two equations when nonlinearity induced self focusing plays important role. 
In order to allow for dispersion and diffraction effects
to develop over larger propagation distance we decreased the peak intensity used in \cite{BK-1997} by a factor of two 
to $2\times 10^{13}\;\mbox{W/cm}^2$
for analyzing propagation in a Kerr medium with $A_{NL}=(2\pi)^{-1}n_0 n_2 |A|^2 A$ where $n_2=3\times 10^{-13}\;$cm$^2$/W is 
the nonlinear index of refraction
and $|A|^2$ is normalized to give the intensity.

\begin{figure}[h]
\centering
\includegraphics[width=70mm]{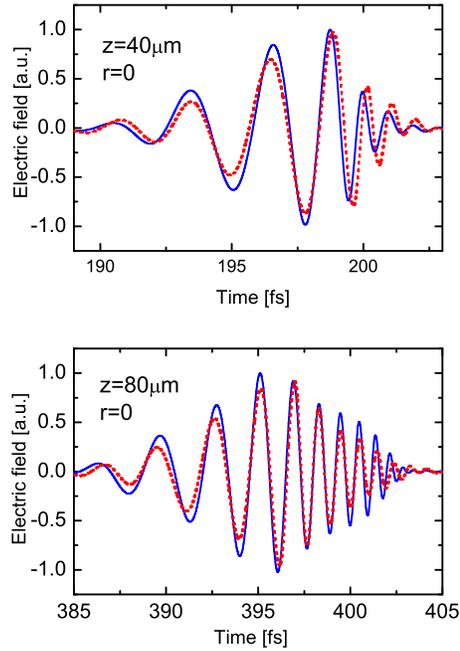}
\caption{Electric field at $z=40\;\mu$m (upper panel) and $z=80\;\mu$m (lower panel) for $r=0$. The full line corresponds to
the cut-off wavelength $\lambda_m=5.3\;\mu$m while the dash line is for $\lambda_m=1.04\;\mu$m which corresponds to the (lower)
frequency at which the envelope takes one half of its maximum value.
\label{fig1}}
\end{figure}

We observe that considering the propagation to be lossless for the above pulse properties and propagation distances 
not exceeding 100$\;\mu$m is a very good approximation since the imaginary part of the refractive index of considered silica
glass in the frequency interval where the spectral distribution of the pulse is not less than 2\% of its maximum is not exceeding
0.0004 \cite{Kitamura-2007} implying an attenuation length larger than 1500$\;\mu$m and even at the cut-off wavelength $\lambda_m=5.3\;\mu$m
(see caption of figure~\ref{fig1}) it is 140$\;\mu$m. In case of strong dispersion or propagation distances exceeding the 
attenuation length more careful treatment is required prefering solution of the full nonlinear wave equation \cite{Palombini-2010}.
 
We start by analyzing propagation of a beam with Gaussian transverse distribution whose amplitude is proportional to
$\exp[-r^2/(2a_r^2)]$ with $a_r=3\;\mu$m. The transverse-derivative term in expression (6) of \cite{BK-1997} contains
the term $\omega+\omega_0$ in the denominator and thus diverges when the envelope spectrum extends close to $-\omega_0$
which can be the case for single-cycle pulses. It is also present in the studied example which means that in order to avoid
overflow during iterations one has to implement a cut-off for $\omega+\omega_0$ values approaching zero. Our simulations
show that this value can be significantly smaller than $\omega_0$ (for example $0.1\omega_0$) without causing numerical problems.
In order to avoid underestimation of the electric field the cut-off should be significantly smaller than the value at which
the envelope falls to one half of its maximum as illustrated by figure~\ref{fig1}.
In that figure we compare results for different cut-off values after propagation distances of $40\;\mu$m and
$80\;\mu$m at central ($r=0$) position. We observe similar relative differences for off-axis positions. For a stronger
focused beam with $a_r=1\;\mu$m the differences are even larger. 
In the following comparisons we use the cut-off value $\lambda_m=5.3\;\mu$m which leaves out only 0.5\% of the envelope
spectral distribution. We remark that in our result (\ref{ueq}) the propagation condition $\beta > k_\perp$ applies and in the 
diverging nonlinear term there is no need to impose constraint on the lower limit of $|\beta-k_\perp|$ due to the integrable nature 
of the singularity. 

\begin{figure}[h]
\centering
\includegraphics[width=120mm]{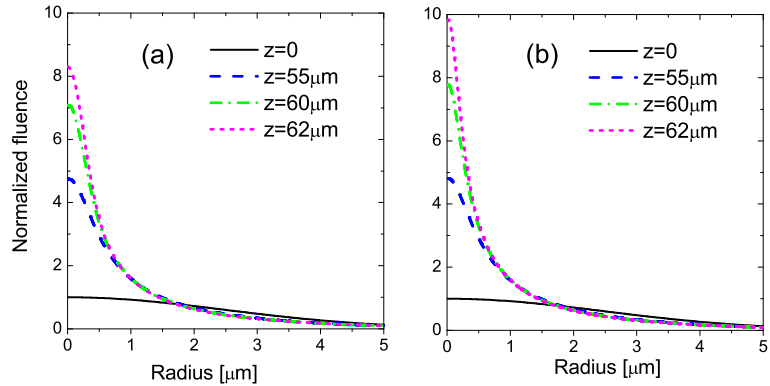}
\caption{Transverse fluence profiles for initial distribution with $a_r=3.5\;\mu$m. 
Our result (a) and using the result from integrating the corresponding equation (6) of \cite{BK-1997} (b).  
\label{fig2}}
\end{figure}

\begin{figure}[h]
\centering
\includegraphics[width=80mm]{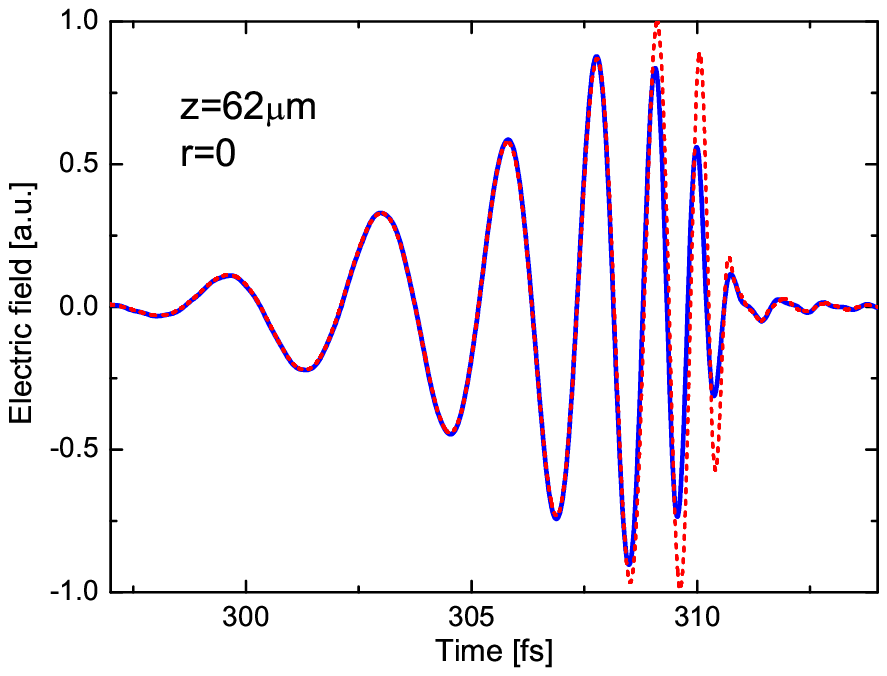}
\caption{Time dependence of the electric field at $r=0$ after propagation of $62\;\mu$m. 
Solid line is our result and dash line is obtained by using the result of the equation from \cite{BK-1997}.  
\label{fig3}}
\end{figure}

Next, we compare the transverse profiles of the pulse by plotting the fluence as function of transverse coordinate 
after different propagation distances using the solution
of (\ref{ueqs}) and of the corresponding expression (6) of  \cite{BK-1997}. We calculate the fluence by 
integrating $|E(z,r,\omega)|^2$ over the frequency $\omega$. 
Figure~\ref{fig2} shows the transverse fluence profiles for different propagation distances starting with a Gaussian
amplitude distribution $\exp[-r^2/(2a_r^2)]$ with $a_r=3.5\;\mu$m. 
In figure~\ref{fig2} we see the effect of self focusing which is dominant if the intensity is sufficiently high and
the focusing of incoming pulse not too strong. The influence of the initial shape of the pulse on the propagation is quite
important similarly to the case when femtosecond pulses propagate in air \cite{Kovachev-2007,Kovachev-2013}.
We remark that in case of studied pronounced self focusing the resultant fluence can be quite high and not far from but
not reaching the damage limit for fused silica \cite{Lenzner-1998}.
In figure~\ref{fig3} we compare the time dependence of the electric field of the pulse 
after propagating $62\;\mu$m at $r=0$ using the two approaches.  
Our results are in general close to the results obtained by using the SEWA approach. However, in case of 
pronounced self focusing corresponding to propagating $z\approx 60\;\mu$m we observe that SEWA leads up to 20\% larger central 
fluence values compared to our result. It is not 
difficult to understand the cause of this behaviour by recalling that SEWA amounts to dropping the transverse component of the
wave vector in the square root of the nonlinear term in equation (\ref{ueq}) thus diminishing the magnitude of that contribution 
for field components with
nonzero transverse coordinate. This in turn leads to relative enhancement of the nonlinear effect for the central region of pulse. 
We can see from figure~\ref{fig3}
that this enhancement comes from larger electric-field amplitude in the descending part of the envelope while the self-phase-modulation
effect does not show noticeable difference. We also observe pronounced difference between the two approaches for large enough values of
the transverse coordinate where the electric field is typically two orders of magnitude smaller than at central position. In this off-axis 
region SEWA typically underestimates the electric-field magnitude.

\begin{figure}[h]
\centering
\includegraphics[width=90mm]{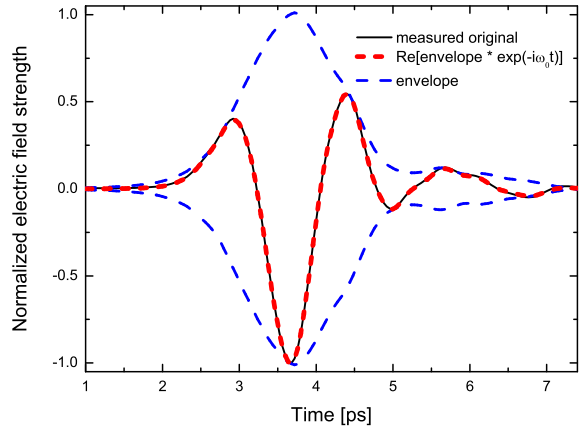}
\caption{The envelope function (dash line) and the measured electric field (solid line) together with the electric 
field calculated using the envelope function.  
\label{fig-env}}
\end{figure}

\begin{figure*}[h]
\includegraphics[width=110mm]{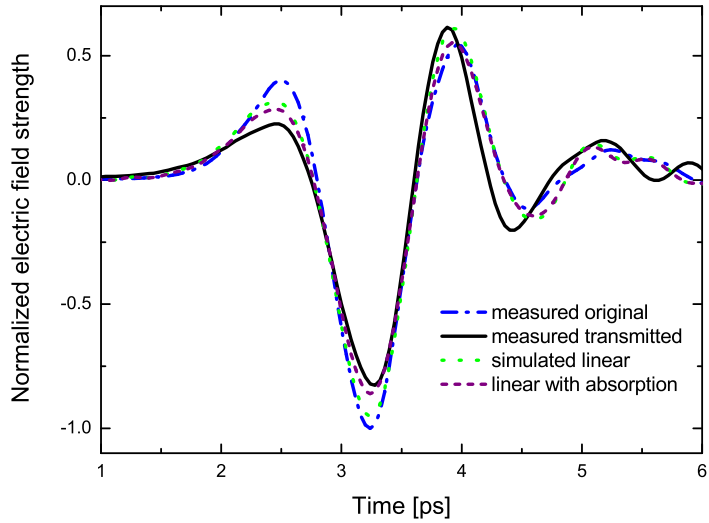}
\centering
\caption{The measured original (dash-dot line), the measured transmitted electric field (solid line) and the calculated transmitted 
field assuming linear propagation with dispersion from \cite{Palfalvi-2005} without absorption (dot line) and with 
absorption coefficient $\alpha=1\;$cm$^{-1}$ (dash line).  
\label{fig-thz1}}
\end{figure*}

We now turn to analyzing terahertz single-cycle pulse propagation as exemplified by observation of nonlinear lattice
response in \cite{Hebling-2009a} where a pulse with peak intensity of 100 MW/cm$^2$ was transmitted through a 
2 mm thick LiNbO$_3$ crystal cooled to 80K. Based on the measured incoming (reference) electric field strength 
we construct a complex envelope function by retaining only the positive-frequency part of the Fourier transform and then 
introducing a shift by the carrier frequency $\omega_0=3.75\;$ps$^{-1}$ obtained using the formula
$\omega_0=\int_0^\infty \omega |E(\omega)|^2 \rmd \omega/ \int_0^\infty |E(\omega)|^2 \rmd \omega$.
The measured reference electric field and the one calculated using the envelope function, together with its shape are shown in 
figure~\ref{fig-env}.
In figure~\ref{fig-thz1} the observed transmitted field strength is shown next to the reference one and the simulated 
result assuming linear propagation with and without absorption and with frequency-dependent refractive index of the form $n=A+b\,f^2+C\,f^4$ with
$A=4.73,\;B=1.5\,10^{-5},\;C=8.5\,10^{-10}$ and $f\equiv \omega/(2\pi c)$ with $\omega/ c$ expressed in cm$^{-1}$ \cite{Palfalvi-2005}.

In this example absorption is not completely negligible but has a small effect as seen from figure~\ref{fig-thz1} and measurements 
reported in \cite{Palfalvi-2005}. Comparing the incoming
and the transmitted pulse we conclude that the absorption coefficient at the carrier frequency should be $\alpha(\omega_0)\approx 1\;$cm$^{-1}$
and taking into account the measured increase with frequency \cite{Palfalvi-2005} we arrive at the value $\alpha(\omega_m)\leq 1.5\;$cm$^{-1}$
where $\omega_m\approx 8\;$ps$^{-1}$ is the frequency at the effective upper limit of the frequency spectrum of the pulse. Since the real part
of the propagation constant $\beta_r(\omega_m)\approx 1200\;$cm$^{-1}$ we ascertain that the condition of weak dispersion is satisfied at 
the high end of spectrum and
also for smaller frequencies characterizing the pulse (measurements extending down to 0.25$\;$THz did not show increase of absorption
coefficient \cite{Unferdorben-2015}). We also observe that the attenuation length is at least three times the propagation distance  
for pulse frequencies.

\begin{figure*}[h]
\centering
\includegraphics[width=110mm]{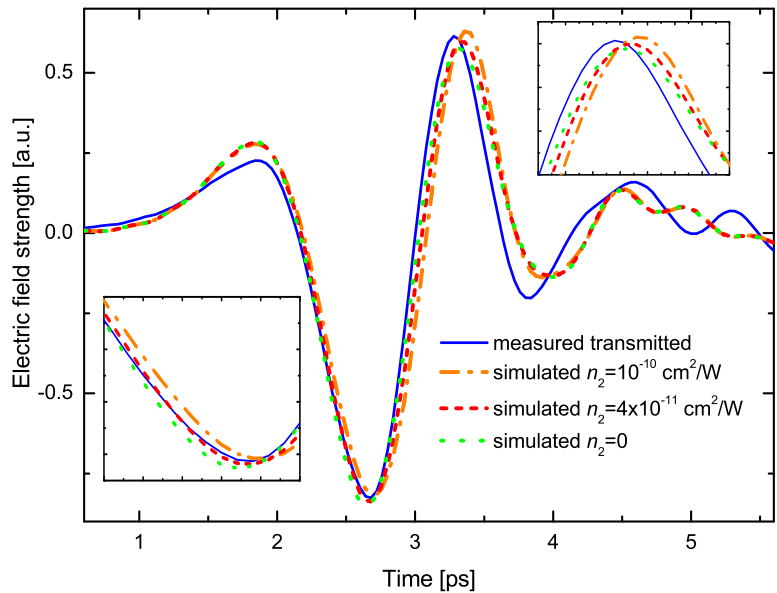}
\caption{Comparison of transmitted and simulated results for different values of nonlinear index of refraction assuming uniform 
transverse distribution.   
\label{fig-thz2}}
\end{figure*}

Introducing a third order Kerr type nonlinearity clearly improves the agreement of the simulated and measured results, notably the
increased distance between the first positive and first negative lobes as can be seen
in figure~\ref{fig-thz2}. In this case we did not consider diffraction effects by assuming a uniform transverse distribution. The value 
of the nonlinear index of refraction giving the closest overall agreement with measurement corresponds to $n_2\approx 4\times 10^{-11}\;$cm$^2$/W 
which is almost four orders of magnitude larger than its value in the visible part of spectrum. 
We also included the second-order nonlinear effect giving rise to the complex nonlinear-polarization envelope \cite{Conforti-2010,Seres-2000} 
\begin{eqnarray}
A_{NL}^{(2)}(z,{\bi{r}}_\perp,t) &=& \frac{d_{33}}{2\pi} \{ A(z,{\bi{r}}_\perp,t)^2 \exp [\rmi (\beta_0 z-\omega_0 t)] \nonumber \\
&+& 2|A(z,{\bi{r}}_\perp,t)|^2 \,\exp [-\rmi (\beta_0 z-\omega_0 t)]\}
\end{eqnarray}
with the restriction that only the positive-frequency part of the polarization should be taken by enforcing the  
constraint $\omega > -\omega_0$ for the envelope.
For the nonlinear optical coefficient $d_{33}$ we use its relationship to the near-infrared refractive index and clamped electro-optic 
coefficient $r_{33}$ (expression (6) of \cite{Hebling-2008}) leading to $d_{33}\approx -160\;$pm/V with $r_{33}=30.8$ \cite{Sutherland-2003}.
The effect of this term is quite small for the considered intensity and propagation distance but nevertheless we observe that it 
brings the simulated electric-field variation slightly closer to the measured signal in the descending region of intensity.

\begin{figure*}[h]
\centering
\includegraphics[width=110mm]{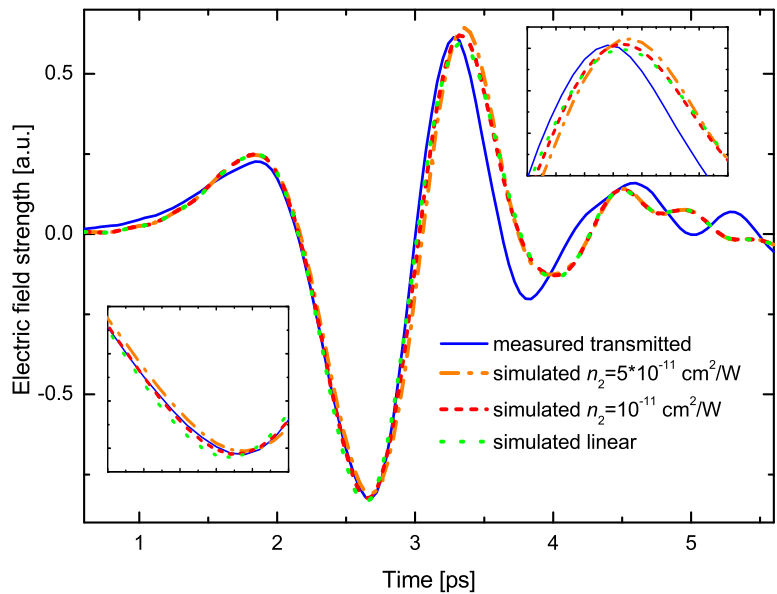}
\caption{Comparison of transmitted and simulated results for different values of nonlinear index of refraction for Gaussian
transverse profile of incoming beam with $a_r=0.6\;$mm.   
\label{fig-thz3}}
\end{figure*}

In order to model diffraction effects we take the incoming pulse with a Gaussian transverse profile and for the measured
values of reference and transmitted field take an average over the width of diameter $D\approx 1\;$mm corresponding to the experimental
condition. Using the estimated value of the width parameter
$a_r\approx 0.6\;$mm shows marked improvement of agreement with measurement for the first oscillation   
and brings the 
preferred value of the nonlinear index of refraction to around $10^{-11}\;\mbox{cm}^2$/W as shown in figure~\ref{fig-thz3}. 
One has to acknowledge that 
the relatively small size of the nonlinear effect for the studied intensity and propagation distance introduces considerable uncertainty 
in that estimate and for more precise determination dedicated experiments with larger intensity and/or propagation distance would be required. 
In case of longer propagation distances surpassing the attenuation length a more careful consideration of attenuative dispersion would
be required as shown by results obtained in \cite{Xiao-1999,Palombini-2010}.
We also investigated the influence of frequency-dependent absorption coefficient slowly increasing with frequency motivated by results of 
 \cite{Palfalvi-2005}
but since it did not produce noticeable change in the agreement with observation we only show results for the constant value 
of $\alpha=1\;$cm$^{-1}$.

\section{Conclusions} 
\label{conc}
We studied diffraction effects during nonlinear propagation of few-cycle light pulses with
axial symmetry in the framework of unidirectional propagation equation derived for the case of weak dispersion by suitably modified
factorization method not relying on the paraxial approximation. The slowly-evolving-wave approximation is
obtained as a special case and numerical simulations show similar results with significant differences present in case
of high intensity accompanied with pronounced self focusing effects. Analysis of short terahertz pulse propagation in LiNbO$_3$ 
shows both nonlinear and diffraction effects despite of short propagation distance and 
indicates presence of a Kerr-type nonlinearity with nonlinear index of refraction $n_2\approx 10^{-11}\;\mbox{cm}^2$/W which is three 
orders of magnitude larger than its value in the visible part of spectrum. This points to significant contribution of lattice 
vibration anharmonicity and indicates pronounced suitability of terahertz pulses for investigating nonlinear lattice dynamics.

\ack

The present work was supported by the Orsz\'agos Tudom\'anyos Kutat\'asi Alapprogramok (OTKA, Hungary) grant K109462 and is 
dedicated to the 650th anniversary of the foundation of University of P\'ecs, Hungary. 

\section*{References}

\bibliographystyle{iopart-num}
\bibliography{paper}

\end{document}